\input{aipcheck}

\documentclass[final,numberedheadings]{aipproc}

\layoutstyle{6x9}

\usepackage{amsfonts}
\usepackage{graphics}

\newcommand{\lb}[1]{\label{#1}}

\newcommand{\bc}{\begin{center}}
\newcommand{\ec}{\end{center}}
\newcommand{\be}{\begin{equation}}
\newcommand{\ee}{\end{equation}}
\newcommand{\bea}{\begin{eqnarray}}
\newcommand{\eea}{\end{eqnarray}}
\newcommand{\ba}[1]{\begin{array}{#1}}
\newcommand{\ea}{\end{array}}

\newcommand{\bt}[1]{\begin{table}[ht]\centering\begin{tabular}{#1}}
\newcommand{\et}[1]{\end{tabular}\caption{\small#1}\end{table}}

\begin{document}


\title{Canonical Functional Quantization of Pseudo-Photons in Planar Systems}
\author{P. Castelo Ferreira}{address={CENTRA, Instituto Superior T\'ecnico, Av. Rovisco Pais, 1049-001 Lisboa, Portugal}}

\begin{abstract}
Extended $U_e(1)\times U_g(1)$ electromagnetism containing both a photon and a pseudo-photon
is introduced at variational level and is justified by the violation of the Bianchi
identities in conceptual systems, either in the presence of magnetic monopoles or
non-regular external fields, not being accounted by the standard Maxwell Lagrangian.
It is carried a dimensional reduction that renders a $U_e(1)\times U_g(1)$ Maxwell-BF
type theory and it is considered canonical functional quantization in planar
systems which may be relevant in Hall systems.
\end{abstract}

\maketitle

The Maxwell equations where derived phenomenologically and unified in 1861~\cite{Maxwell}.
At variational level these equations are described by the well known Maxwell Lagrangian
and respective electromagnetic field definitions
\be
{\mathcal{L}}_{\mathrm{Maxwell}}=-\frac{1}{4}F_{\mu\nu}F^{\mu\nu}+A_\nu J^\nu_e\ \ ,\ \ E^i=F^{0i}\ \ ,\ \ \ B^{i}=\frac{1}{2}\epsilon^{ijk}F_{jk}\ ,
\lb{L_Maxwell}
\ee
The equations of motion are obtained by considering a functional derivation of the action with
respect to the gauge field $A_\mu$
\be
\frac{\delta {\mathcal{L}}_{\mathrm{Maxwell}}}{\delta A_\mu}=0\ \Leftrightarrow\ \partial_\mu F^{\mu\nu}=J_e^\nu
 \Rightarrow \left\{\ba{rcl}\displaystyle\bf{\nabla}\cdot\bf{E}&=&\displaystyle\rho_e\\[3mm] \displaystyle\bf{\nabla}\times\bf{B}-\dot{\bf{E}}&=&\displaystyle{\bf{j}}_e\ea\right.
\lb{EOM_Maxwell}
\ee
which only render half of the Maxwell equations. The remaining equations are obtained by
demanding (or imposing) regularity of the gauge fields, i.e. the Bianchi identities
\be
\epsilon^{\mu\nu\lambda\rho}\partial_\nu F_{\lambda\rho}=0\,\Rightarrow\, \left\{\ba{r}\displaystyle\bf{\nabla}\cdot\bf{B}= 0\\[3mm] 
			                         \displaystyle\bf{\nabla}\times\bf{E}+\dot{\bf{B}}= 0\ea\right.
\lb{BI}
\ee
The Maxwell Lagrangian~(\ref{L_Maxwell}) is a successful functional principle in most fields of physics dealing with electromagnetic interaction, both at classical and quantum level. It was the first example of fundamental interactions
unification based in the phenomenologically derived Maxwell equations and it is in the basis of todays particle physics, in particular quantum field theory and the standard model. However does not reproduce the Bianchi Identities which
are imposed externally. Although for most of physical systems the Maxwell Lagrangian is enough to achieve a complete
description at variational level the last point clearly holds a problem when trying to describe some conceptual
systems. Examples of such systems are the description at variational level of magnetic monopoles~\cite{Dirac,WY,CF,Sing,EB}
through the inclusion of magnetic currents $J_g^\mu=(\rho_g,j_g^i)$ and external non-regular fields~\cite{pseudo} which imply:
\begin{itemize}
\item violation of the Bianchi identities~(\ref{BI});
\item are not deducible from the Maxwell Lagrangian~(\ref{L_Maxwell});
\item for monopoles we have extended singularities, either the Dirac string~\cite{Dirac} or the Wu-Yang fiber bundle~\cite{WY}.
\end{itemize}
So we conclude that for both these cases the Maxwell Lagrangian~(\ref{L_Maxwell}) is an incomplete description
of the electromagnetic interactions. We recall that the existence of Magnetic monopoles are the only theoretical
justification for the quantization of electric charge~\cite{Dirac}, however have so far not been experimentally
detected and due to Dirac quantization condition $eg=2\pi\hbar n$ are in a strong coupling regime hence confined.
As for external non-regular fields seems to be relevant only in systems where either flux-tubes (strings in $3+1$-dimensions)~\cite{flux}
or vortexes (in $2+1$-dimensions) are present~\cite{fqhe}.

A possible solution for this problem is to consider an extended
Abelian gauge group $U_e(1)\times U_g(1)$, i.e. in addition to the standard
\textit{electric} vector gauge field (photon) to consider a \textit{magnetic}
pseudo-vector gauge field (pseudo-photon). Let us take the following assumptions:
\begin{enumerate}
\item Parity ($P$) and Time-inversion ($T$) invariance of electromagnetic interactions;
\item existence of only one electric and one magnetic physical fields.
\end{enumerate}
Both this assumptions are justified experimentally. Given the above assumptions,
up to a sign choice of the topological term, we get the only possible Lagrangian~\cite{CF,Sing,EB}
\be
{\mathcal{L}}_4=\displaystyle \frac{1}{4}\left[-F_{\mu\nu}F^{\mu\nu}+G_{\mu\nu}G^{\mu\nu}+\epsilon^{\mu\nu\lambda\rho}F_{\mu\nu}G_{\lambda\rho}
+e\,\left(A_\mu-\tilde{C}_\mu\right)J_e^\mu-g\,\left(\tilde{A}_\mu-C_\mu\right)J_g^\mu\right]
\lb{L_AC}
\ee
with the electromagnetic field definitions
\be
E^i=F^{0i}-\frac{1}{2}\epsilon^{0ijk}G_{jk}\ \ ,\ \ \ B^i=G^{0i}+\frac{1}{2}\epsilon^{0ijk}F_{jk}\ .
\lb{EB_fields}
\ee
For further details in this derivation and definitions of $\tilde{A}$ and $\tilde{C}$ in terms of
the connection Hodge duals $\tilde{F}$ and $\tilde{G}$ see~\cite{EB}. In addition we note that the
coupling constants $e$ and $g$ correspond to the electric and magnetic unit charges and obey Dirac's condition.
Both in the presence of magnetic monopoles and non-regular external electromagnetic fields the equations of motion for the $A$ and $C$ field now reproduce the full Maxwell equations in terms of the above electromagnetic field definitions~(\ref{EB_fields}).

Next let us consider a dimensional reduction of extended $U_e(1)\times U_g(1)$ electromagnetism
with action given by~(\ref{L_AC}). We are considering a planar system with a finite thickness $\delta_\perp$
and the coordinate index conventions as pictured in figure~\ref{fig} and as derived in~\cite{dim_red}.
 \begin{figure}[h]
  \includegraphics[height=2cm]{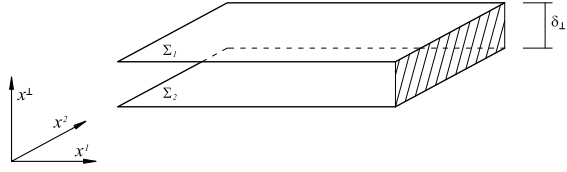}
  \caption{The planar system of finite thickness $\delta_\perp$ with boundaries $\Sigma_1$ and $\Sigma_2$
   and respective \hspace{2cm} $1+(2+1)$ coordinate decomposition $x^I=x^\mu,x^\perp$ with $I=\mu,\perp$, $\mu=0,i$ and $i=1,2$.\lb{fig}}
\end{figure}
Taking a $1+(2+1)$-dimensional decomposition of the Lagrangian~(\ref{L_AC}) assuming gauge field localization in the planar system for the range $\displaystyle x^\perp\in\left[-\delta_\perp/2,+\delta_\perp/2\right]$, Neumann boundary conditions for the gauge fields $\displaystyle A_\perp=C_\perp=0$ which also implies that the gauge is partially fixed
along the orthogonal direction $\displaystyle\partial_\perp\Lambda=0$, constant gauge fields along the
orthogonal direction $\displaystyle\partial_\perp A_\mu=\partial_\perp C_\mu=0$, that only electric 4-current are present, that the gauge fields are regular and that both boundaries are identified as $\displaystyle\Sigma(x^\perp=0)\cong\Sigma_1\cong\Sigma_2$ such that the topological boundary action
terms read~\cite{Witten} $\displaystyle\int_{\Sigma_1-\Sigma_2}\epsilon^{\mu\nu\lambda}A_\mu\partial_\nu C_\lambda\equiv \frac{k}{2}\,\int_\Sigma\epsilon^{\mu\nu\lambda}A_\mu\partial_\nu C_\lambda$ (with $k=0,\pm 1$) and
integrating the orthogonal coordinate $S_3=\int dx^3 {\mathcal{L}}_3=\int dx^3 dx^\perp{\mathcal{L}}_4$ we obtain
\be
{\mathcal{L}}_3=\frac{1}{4}\left[-\delta_\perp F_{\mu\nu}F^{\mu\nu}+\delta_\perp G_{\mu\nu}G^{\mu\nu}+4\delta_\perp A_\mu J_e^\mu+\frac{k}{2}\epsilon^{\mu\nu\lambda}A_{\mu}G_{\nu\lambda}+\frac{k}{2}\epsilon^{\mu\nu\lambda}C_{\mu}F_{\nu\lambda}\right]\ ,
\lb{L_3}
\ee
with the electromagnetic field definitions in the planar system given by 
\be
E^i=F^{0i}\ \ ,\ \  E^\perp=-G_{12}\ \ ,\ \  B^\perp=F_{12}\ \ ,\ \ B^i=-G^{0i}\ ,
\ee
obtained from the definitions~(\ref{EB_fields}).
We note that the orthogonal component of the electric field and the longitudinal component
of the magnetic field are described by the theory in terms of the pseudo-photon. This is a
completely new feature not present in the standard $U(1)$ Maxwell or Maxwell Chern-Simons theories
in planar systems.

This theory has been applied to the fractional Hall effect in~\cite{fqhe} where it is considered
an effective Landau-Ginzburg Chern-Simons model for anyons with the dynamical gauge field being
the pseudo-photon field $C$ and with coupling constants $e\alpha$ and $g\hat{\beta}$ given
in terms of a scalar $\alpha$ and pseudo-scalar $\hat{\beta}$ standing respectively for the
average number of unit flux electric and magnetic vortexes per electron such that the equations
of motion hold an orthogonal induced electric field and the Hall conductance
\be
E^\perp=-\frac{e\alpha}{g\hat{\beta}}B\ \ ,\ \ 
j^i=\frac{e\alpha}{2g\hat{\beta}}\epsilon^{0ij}E_j=\hat{\sigma}_H\epsilon^{0ij}E_j\ .
\ee
Both these expressions are invariant under $P$ and $T$ discrete symmetries (as opposed
to the standard Chern-Simons models that describe the Hall effect)
and the $\Phi_0$ quantization is in this framework due to Dirac's condition
such that $\hat{\sigma}_H=\frac{e\alpha}{2g\hat{\beta}}=\frac{e}{\Phi_0}\,\frac{\alpha}{\hat{\beta}}$.
Also the model gives a theoretical justification for the experimentally measured fractional charge $e^*=\frac{1}{2n+1}$ for every filling fraction $\nu=\frac{p}{2n+1}$ independently of $p$, gives a theoretical justification
at macroscopical level for the low energy contribution to Laughlin's wave function solutions due to the negative
energy contributions of pseudo-photon excitations (which are ghost or phantoms) and the orthogonal electric
potential (experimentally measured) is due to pseudo-photon electric vortexes which may explain the
stronger inter-layer correlations, hence the existence of BEC condensates in bi-layer electron-electron
Hall systems instead of its existence in electron-hole Hall systems as originally expected~\cite{nature}.

Taking a functional quantization procedure the canonical momenta are
\be
\pi_A^i=-\delta_\perp\,F^{0i}+\frac{1}{4}\epsilon^{ij}C_j=-i\frac{\delta}{\delta A_i}\ \ \ ,\ \ \ \pi_C^i=+\delta_\perp\,G^{0i}+\frac{1}{4}\epsilon^{ij}A_j=+i\frac{\delta}{\delta C_i}\ .
\ee
We are considering a functional Schro\"odinger picture and the difference of sign in the functional
derivatives is due to the respective difference of signs in the kinetic term for the fields $A$ and
$C$ in the Lagrangian~(\ref{L_3})~\cite{Govaerts}). For $k=+1$ and from the usual definition of the
Hamiltonian ${\mathcal{H}}_{AC}=+\pi^i_A\partial_0A_i+\pi^i_C\partial_0C_i-{\mathcal{L}}_{C}$
we obtain three functional constraints for the ground state wave functional $\Phi_{(0,0)}[A,C]$~\cite{dim_red},
the Hamiltonian constraint ${\mathcal{H}}\Phi_{(0,0)}[A,C]={\mathcal{E}}_{(0,0)}\Phi_{(0,0)}[A,C]$ and
the two Gauss' laws ${\mathcal{G}}_A\Phi_{(0,0)}[A,C]=0$ and ${\mathcal{G}}_A\Phi_{(0,0)}[A,C]=0$
for the operators
\be
\ba{rcl}
\displaystyle{\mathcal{H}}&=&\displaystyle+\frac{1}{2}\left(i\frac{\delta}{\delta A_i}-\frac{1}{4}\epsilon^{ij}C_j\right)^2+\frac{\delta_\perp}{4}F_{ij}F^{ij}-\frac{1}{2}\left(i\frac{\delta}{\delta C_i}+\frac{1}{4}\epsilon^{ij'}C_{j'}\right)^2-\frac{\delta_\perp}{4}G_{ij}G^{ij},\\[5mm]
\displaystyle{\mathcal{G}}_A&=&\displaystyle\partial_i\left(i\frac{\delta}{\delta A_i}-\frac{1}{4}\epsilon^{ij}C_j\right)\ \  \ ,\ \ \ 
{\mathcal{G}}_C=\partial_i\left(i\frac{\delta}{\delta C_i}+\frac{1}{4}\epsilon^{ij}A_j\right)\ .
\ea
\lb{constraints}
\ee
The simplest solution for both Gauss' laws is the topological ground state solution
\be
\Phi_{(0,0)}[A,C]=e^{-i\frac{1}{4}\int dx^2\epsilon^{ij}A_iC_j}\ ,
\ee
where the two labels stand for the state numbers corresponding to photon and pseudo-photon
fields. Replacing this solution in the Hamiltonian constraint we obtain
the functional energy
\be
{\mathcal{E}}_{(0,0)}=\frac{1}{4}\int dx^2\left(\delta_\perp F^{ij}F^{ij}-\frac{1}{2}\,A^iA^i-\delta_\perp G^{ij}G^{ij}+\frac{1}{2}\,C^iC^i\right)\ ,
\ee
which is minimized for the solutions of the functional equations
\be
\delta_\perp\int dx^2\,F_{ij}F^{ij}=\frac{1}{2}\int dx^2\,A_i A^i\ \ ,\ \ \delta_\perp\int dx^2\,G_{ij}G^{ij}=\frac{1}{2}\int dx^2\,C_iC^i\ ,
\ee
which hold either the trivial solutions or the vortex solutions
\be
A^i=\pm\frac{\epsilon^{ij}(r_j-\bar{r}_j)}{|r-\bar{r}|}\ \ ,\ \ C^i=\pm\frac{\epsilon^{ij}(r_j-\bar{r}_j)}{|r-\bar{r}|}\ \ ,\ \ R=\sqrt{2\pi}\delta_\perp\ \ ,\ \ \delta_\perp=\frac{l_m}{\sqrt{\pi}}.
\ee
The finite vortex radius $R$ is required for finiteness of the vortex energy and charge.
Here the relation between the planar system thickness $\delta_\perp$ and the magnetic length $l_m$
is not derived from the theory, instead we are imposing it using a direct analogy with the
known physical results, in particular in Hall systems.
As considered in~\cite{Dbranes} by considering the lowering and raising operators
${\hat{E}}_z= -i\frac{\delta}{\delta A_z}+\frac{ik}{4}C_{\bar{z}}$ , ${\hat{E}}_{\bar{z}}= -i\frac{\delta}{\delta A_{\bar{z}}}-\frac{ik}{4}C_z$, ${\hat{B}}_z= +i\frac{\delta}{\delta C_z}+\frac{ik}{4}A_{\bar{z}}$ and ${\hat{B}}_{\bar{z}}= +i\frac{\delta}{\delta C_{\bar{z}}}-\frac{ik}{4}A_z$
we can build the excited wave functionals
$\Phi_{(n,m)}=\left({\hat{E}}_{\bar{z}}\right)^n\left({\hat{B}}_{\bar{z}}\right)^m\Phi_{(0,0)}[A,C]$
which clearly seems consistent with Laughlin's wave functions although it is still missing a direct
proof (work in progress). Also considering the field decomposition $A_\mu=A_\mu^{\mathrm{ext}}+a_\mu$
into internal fields $a$ and external fields $A^{\mathrm{ext}}$ the wave functional solution is a
correction to $\Phi_{(0,0)}$
\be
\Psi_A[a,C]=e^{-i\frac{k}{4\delta_\perp}\int dx^2\epsilon^{ij}\left(A^{\mathrm{ext}}_i+a_i\right)C_j}=e^{-i\frac{k}{4\delta_\perp}\int dx^2\epsilon^{ij}A^{\mathrm{ext}}_iC_j}\Phi_{(0,0)}.
\ee
Considering a measure-shift in the path integral $a\to a-A^{\mathrm{ext}}$ and integrating
the $a$ field we conclude that the path integral (the functional partition function) is equivalently
formulated in terms of the wave functionals for the $C$ field only given by
$\Psi_A[C]=\left(1-ik/(4\delta_\perp)\epsilon^{ij}A^{\mathrm{ext}}_iC_j+\ldots\right)\Phi_{0}[C]$.

Hence these results are a preliminary prove of the simultaneous existence of both electric and
magnetic vortexes in the theory  due, respectively, to the pseudo-photon and photon field, as well
as a justification for the internal photon $a$ being, in the macroscopical model for the Hall effect,
a non-dynamical field. However it is still missing a prove of fractional statistics for anyons
which should be achieve when non-trivial homology is considered~\cite{Semenoff}. This is obtained by
considering the effect of electrons in the system corresponding to charge vertex insertions in the plane,
hence creating non-trivial homology cycles~\cite{Dbranes}.\\[-2mm]

\noindent[This work was supported by SFRH/BPD/17683/2004 and SFRH/BPD/34566/2007].\\[-6mm]

\end{document}